# Io's Optical Aurorae in Jupiter's Shadow

Carl Schmidt[1,2], Mikhail Sharov[3], Katherine de Kleer[4], Nick Schneider[5], Imke de Pater[6], Phillip H. Phipps[7,8],
Albert Conrad[9], Luke Moore[1,2], Paul Withers[1,2], John Spencer[10], Jeff Morgenthaler[11], Ilya Ilyin[12],
Klaus Strassmeier[12], Christian Veillet[9], John Hill[9], and Mike Brown[4]
[1] Department of Astronomy, Boston University, USA
[2] Center for Space Physics, Boston University, USA
[3] Department of Physics and Astronomy, University of Texas at San Antonio, USA
[4] Division of Geological and Planetary Sciences, California Institute of Technology, USA
[5] Laboratory for Atmospheric and Space Physics, University of Colorado at Boulder, USA
[6] Earth and Planetary Science, University of California, Berkeley, USA
[7] University of Maryland, Baltimore County, USA
[8] NASA Goddard Space Flight Center, USA
[9] Large Binocular Telescope, University of Arizona, USA
[10] Southwest Research Institute, Boulder, CO, USA
[11] Planetary Science Institute, USA
[12] Leibniz-Institute for Astrophysics Potsdam, Germany
*Received 2022 May 16; revised 2022 July 21; accepted 2022 July 30; published 2023 February 16*

## Abstract

Decline and recovery timescales surrounding eclipse are indicative of the controlling physical processes in Io's atmosphere. Recent studies have established that the majority of Io's molecular atmosphere, $SO_2$ and SO, condenses during its passage through Jupiter's shadow. The eclipse response of Io's atomic atmosphere is less certain, having been characterized solely by ultraviolet aurorae. Here we explore the response of optical aurorae for the first time. We find oxygen to be indifferent to the changing illumination, with [O I] brightness merely tracking the plasma density at Io's position in the torus. In shadow, line ratios confirm sparse $SO_2$ coverage relative to O, since their collisions would otherwise quench the emission. Io's sodium aurora mostly disappears in eclipse and e-folding timescales, for decline and recovery differ sharply: ∼10 minutes at ingress and nearly 2 hr at egress. Only ion chemistry can produce such a disparity; Io's molecular ionosphere is weaker at egress due to rapid recombination. Interruption of a $NaCl^+$ photochemical pathway best explains Na behavior surrounding eclipse, implying that the role of electron impact ionization is minor relative to photons. Auroral emission is also evident from potassium, confirming K as the major source of far red emissions seen with spacecraft imaging at Jupiter. In all cases, direct electron impact on atomic gas is sufficient to explain the brightness without invoking significant dissociative excitation of molecules. Surprisingly, the nonresponse of O and rapid depletion of Na is opposite the temporal behavior of their $SO_2$ and NaCl parent molecules during Io's eclipse phase.

*Unified Astronomy Thesaurus concepts:* Galilean satellites (627); Eclipses (442); High resolution spectroscopy (2096); Aurorae (2192); Planetary magnetospheres (997)

## 1. Introduction

Each orbit, as Io passes through Jupiter's shadow, its surface cools. Sublimation support of Io's $SO_2$ atmosphere is sensitive to small changes in its surface temperature, and the bulk atmosphere rapidly condenses. This potentially leaves direct volcanic outgassing as the primary mechanism sustaining Io's atmosphere in shadow. Tsang et al. (2016) estimated that the global $SO_2$ column decreases from 2.0–2.5 × $10^{16}$ to ∼0.5 × $10^{16}$ cm$^{-2}$ within 40 minutes of Io's ingress. De Pater et al. (2020) showed that the $SO_2$ column density in localized regions can remain stable throughout eclipse, but the fractional coverage of $SO_2$ over Io's surface is decreased by a factor of 2–3, broadly consistent with the Tsang et al. spatially unresolved measurement. Though an assumption of the atmospheric temperature is necessary, both mid-infrared and millimeter measurements are straightforward proxies for the changing $SO_2$ column abundance as Io passes through Jupiter's shadow.

Photolysis and plasma bombardment also break down Io's molecular atmosphere into an atomic form. A similar depletion of the atomic atmosphere in eclipse would imply that the dissociation of molecules into atoms proceeds rapidly. While the plasma torus bombardment remains continuous, eclipses by Jupiter interrupt photochemistry, so the eclipse response can offer insight into photochemical production and loss mechanisms and their timescales.

In eclipse, Io's atomic atmosphere is seen as aurorae. Most observations have been made at far-ultraviolet (FUV) wavelengths, where aurorae contrast against Io's low surface reflectivity, making it easy to compare sunlit and shadowed data. Hubble Space Telescope (HST) measurements show O and S brightness a factor of 2 or 3 weaker in shadow (Clarke et al. 1994; Roth et al. 2014), as expected from rapid dissociation of condensing $SO_2$. New Horizons flyby measurements show decreasing brightness upon ingress for a dayside viewing geometry and little change for a similar ingress light curve viewing the nightside, while egress observations showed a pronounced brightening (Retherford et al. 2007).

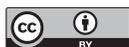

Aurorae are less straightforward to interpret compared to $SO_2$ measurements. Emission can originate from electron impact, radiative dissociation of molecules, or dissociative recombination of molecular ions. Each depends not just on the atmospheric abundance but also on the plasma density and temperature, penetration depth, and emission opacity. Aurorae strongly depend on Io's electrodynamic interaction with the plasma torus, but the Io–torus interaction itself could quickly change in eclipse; to some extent, photoionization supplies Io's ionosphere, so recombination and changing sublimation rates will alter Io's ionospheric conductance relative to the Alfvén conductance in the torus. Saur & Strobel (2004) showed that this nonlinear behavior can actually brighten aurorae in eclipse, and that this conductivity ratio, which regulates currents in the Io–Jupiter circuit, is sensitive to volcanic activity. Consensus has not yet emerged on the direct volcanic contribution to Io's atmosphere and its variability. The 1%–3% volcanic contribution to the dayside atmosphere inferred from simulating UV aurorae (Retherford et al. 2007) is well below the 30%–50% inferred from molecular observations (Tsang et al. 2016; de Pater et al. 2020). Nonresponses of the $SO_2$ column have also been reported postegress, suggesting that volcanism, not sublimation, could be its primary supply at times (Tsang et al. 2015). With this confluence of factors, it is challenging to understand how Io's atomic abundance varies as it passes through Jupiter's shadow.

Io also produces optical and near-IR aurorae in eclipse. Galileo's broadband imagery showed that these distributions are similar to FUV emissions, with the brightest spots near the tangent points of the Jovian magnetic field and diffuse glow at the limb (Geissler et al. 1999). Spectroscopy with the Keck High Resolution Echelle Spectrometer (HIRES) instrument identified Galileo's emission sources as [O I] 5577 Å (green line), 6300 and 6364 Å (red line doublet), and the sodium D doublet at 5890 and 5896 Å, with a brightness of 1–25 kR over Io's disk (Bouchez et al. 2000). This comprises the sole ground-based measurement in the optical, but HST WFPC2 spatially resolved the [O I] 6300 Å aurora and found that it glows brightest in Io's wake region, an effect attributed partly to viewing geometry but also to a locally enhanced electron population (Moore et al. 2010). Cassini also obtained imagery of Io in eclipse during its passage through the Jovian system. This yielded additional emissions in the 7300–8000 Å range for which potassium aurorae seem the best explanation and also gave marginal detections of emitters between 8000 and 10250 Å, which could be cascades from O- and S-atom Rydberg levels produced by $SO_2$ dissociation (Geissler et al. 2004). Broadband images from New Horizons' LORRI lacked wavelength information but nicely showed that the gas component of the Tvashtar plume produced optical aurorae at high altitudes of 350 km from Io's limb (Spencer et al. 2007; Roth et al. 2011).

The aim of this paper is to extend the ultraviolet studies of Io's eclipse response to wavelengths accessible from the ground. Previous optical and near-IR observations have not explored the potential relationships between brightness and duration in shadow, nor have they characterized modulations with magnetic geometry (though it is seen that the north/south limb of Io that faces the torus is brighter; Geissler et al. 2001; Retherford et al. 2003). However, sunlit emissions indicate that Na escaping Io must be sensitive to the eclipse phase. Grava et al. (2014) found that, postproxy, Na emission at distances >3 Io radii drops by a factor of 4 relative to preeclipse values, and that the recovery time to preeclipse values following egress was >5 hr. This motivates our study, which is the first to characterize how Io's optical aurorae evolve with time in shadow. Spectroscopic observations are described in Section 2. In Section 3, we show the resultant emissions and describe aspects that govern their brightness in shadow. The plausible emission mechanisms are discussed in Section 4. Section 5 shows sunlit sodium emission following egress and describes some of the potentially associated photochemical reactions. Discussion, conclusions, and future work appear in Section 6.

## 2. Observations and Analysis to Isolate Io's Emission

Geometries near quadrature between Jupiter, the Sun, and Earth are required to observe Io in shadow while also maximizing the elongation from Jupiter's bright limb. Eclipse observations are challenging compared to sunlit measurements in that they have no light source for closed-loop guiding and require blind nonsidereal tracking. Tracking sunlit satellites using the time-dependent nonsidereal rates informed the accuracy of open-loop blind tracking and generally showed appreciable drift of an arcsecond or more within an hour due to, e.g., telescope flexure or inaccuracies of the refraction model. Repointing the telescope between exposures ultimately proved a more reliable technique than continuous nonsidereal tracking but at the cost of less integration time on Io. Offsets between adjacent satellites in sunlight were used to characterize the accuracy of a blind offset to Io in shadow. This procedure proved reliable, but only with attention to the precise timing of each satellite's ephemeris and the directionality of the offset to avoid backlash in the drive gears. Table 1 lists the observations obtained using this technique.

High spectral dispersion is needed to isolate Io's atomic lines from the bright continuum of scattered light near Jupiter, as well as from Earth's emission lines at the same wavelengths, from which Io's are Doppler shifted. The data in this study were obtained by three high-resolution optical spectrographs. The primary instrument was the cross-dispersed ARC Echelle Spectrograph (ARCES) on the ARC 3.5 m telescope at Apache Point Observatory (APO; Wang et al. 2003). All ARCES observations used its $1''\!.6 \times 3''\!.2$ slit, which corresponds to a resolving power of $R \sim 31{,}500$. Figure 1 shows an example of the planet–satellite geometry for the first night of observations. The ARCES guide camera ran continuously during all observations, allowing for a pointing confirmation using Jupiter's limb and other satellites, per panels (c) and (d). HIRES, on the Keck I 10 m telescope (Vogt et al. 1994), was also used. Its $7''\!.0 \times 1''\!.722$ slit was oriented tangent to Jupiter's limb, obtaining $R \sim 24{,}700$ spectra with Io and scattered Jovian light sampled within the same aperture. Continuous tracking with Keck I also exhibited significant drift, so it was necessary to realign the pointing using Ganymede between each exposure of Io in eclipse. The Potsdam Echelle Polarimetric and Spectroscopic Instrument (PEPSI) is a fiber-fed spectrograph at the $2 \times 8.4$ m Large Binocular Telescope (LBT; Strassmeier et al. 2015, 2018). We used the $2''\!.3$ diameter aperture mode for $R \sim 50{,}000$ (Figure 2). PEPSI's exposures alternated cross-disperser settings to target different wavelength ranges, while ARCES and HIRES data span the optical wavelength range nearly continuously, with occasional gaps between orders in the near-IR.

Table 1
Spectroscopic Measurements of Io's Atmosphere Surrounding Eclipse

| UT Date | Time | Telescope/Instrument | Orbital Longitudes (deg) | Ingress or Egress | No. of Frames Sunlit/Penumbra/Umbra |
|---|---|---|---|---|---|
| 2018-03-20[a] | 10:33–11:24 | APO/ARCES | 349–356 | I | 6/1/6 |
| 2018-08-07 | 07:26–08:29 | Keck/HIRES | 359–8 | E | 0/0/10 |
| 2019-04-24 | 09:09–11:10 | LBT/PEPSI | 341–358 | I | 3/0/4 |
| 2019-08-12 | 04:56–06:27 | APO/ARCES | 0–13 | E | 4/1/8 |
| 2020-08-23 | 04:49–05:53 | APO/ARCES | 4–13 | E | 4/0/5 |
| 2020-09-08 | 02:47–04:44 | APO/ARCES | 1–18 | E | 6/1/3 |
| 2020-10-01 | 03:04–05:27 | APO/ARCES | 1–21 | E | 8/1/7 |
| 2020-10-17 | 02:01–02:52 | APO/ARCES | 7–14 | E | 4/2/2 |
| 2021-06-09 | 09:25–09:40 | APO/ARCES | 349–352 | I | 1/0/1 |
| 2021-10-29 | 00:58–04:51 | APO/ARCES | 1–34 | E | 24/1/7 |

**Note.**
[a] Concurrent with ALMA molecular observations (de Pater et al. 2020).

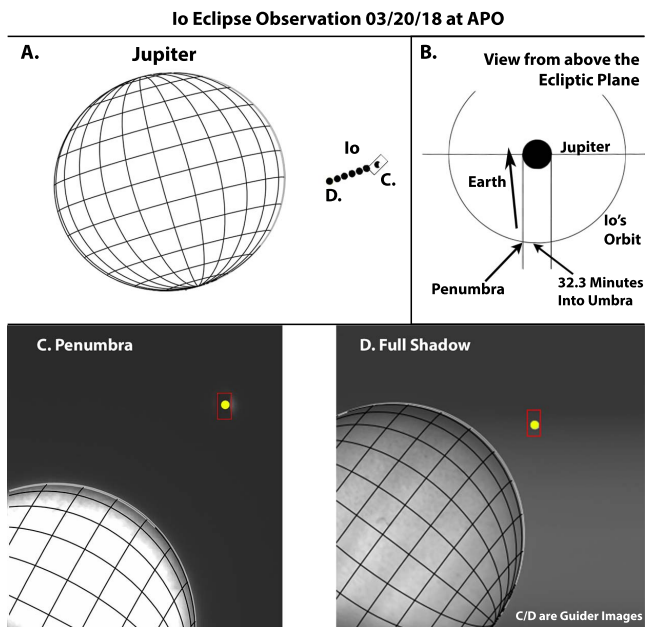

**Figure 1.** Diagram and guider frame images showing the ARCES aperture (red box) during ingress on 2018 March 20.

Even in ideal observing geometry, the optical spectra of Io in Jovian eclipse are strongly contaminated by Jupiter's scattered light. Our initial attempts to remove this scattered light applied spectra of Jupiter's disk that were smoothed, wavelength-shifted, fit, and subtracted from the eclipsed data in order to isolate faint emissions from Io in the residual. This technique seemed to work well, but it produced artifacts upon close inspection. Unlike the spectrum of Jupiter's disk center, which we use for flux calibration, scattered light above the Jovian limb has asymmetric Fraunhofer features; Jupiter's rapid rotation produces a Doppler shift across its disk, and in the sky proximal to Jupiter, scattered light does not derive from all portions of the disk uniformly. The Fraunhofer features are deeper on the red wing at ingress and the blue wing at egress because scattered light originates predominantly from Jupiter's receding and approaching limb, respectively. The results were misleading when we did not account for this initially. The Fraunhofer artifacts were not only characteristic of an emission line profile in shape but also appeared very near Io's Doppler shift by coincidence that is inherent to this observing geometry.

Hence, we caution observers that fully mitigating Jovian scattered light requires a more careful treatment, as we have done in this paper.

Long-slit spectroscopy to sample the Jovian background concurrently is ideal. With HIRES, bright [O I] 6300 Å was used to determine Io's location along the slit, and scattered light was interpolated from spectra at adjacent spatial bins. With PEPSI, sky fibers were used to measure scattered light proximal to the target aperture. With ARCES, it was necessary to obtain dedicated measurements of the scatter at various distances from the Jovian limb. In both of these latter cases, the sky spectrum was paired with the eclipse spectra by cross-correlation on a frame-by-frame basis.

Once the Jovian scattered-light background is subtracted, a residual spectrum from the Earth and Io remains, which can be separated by Doppler shift. Gaussians are fit to Io's line-spread functions and integrated to give the total brightness. Jupiter's reflectance spectrum is applied to calibrate absolute flux (e.g., de Kleer & Brown 2018; Schmidt et al. 2018). A correction is made for the fact that Jupiter fills the slit aperture while Io's filling factor varies, and also for the broadband extinction due to the differential airmass. Emission in this work is reported in Rayleigh units assuming an Io-sized source. The associated column density estimates are disk-averaged, but it warrants reiteration that spatially resolved imaging shows that Io's aurora is highly structured locally.

### 3. Io's Optical Emission in Eclipse

As Bouchez et al. (2000) showed, Io's brightest optical emissions in eclipse are the [O I] red line doublet at 6300 and 6364 Å and the Na D line doublet at 5890 and 5896 Å. Figures 3 and 4 show each following ingress on 2020 March 18 and 2019 April 24. The O and Na aurorae exhibit variability on timescales of minutes. Both weaken following ingress on 2018 March 20, while on 2019 April 24, Na weakens, with O brightening concurrently. The example on these dates demonstrates how the response of sodium and oxygen emissions to eclipse is decoupled and distinct. This behavior points toward a relative change between their neutral columns, since it cannot be attributed to the above-described ambiguity in the changing plasma excitation rates alone.

The disk-averaged sodium and oxygen brightness is seen as a time series in Figure 5. During the 2018 March 20 eclipse, concurrent measurements were also made at the Atacama Large Millimeter/submillimeter Array (ALMA; de Pater et al. 2020).

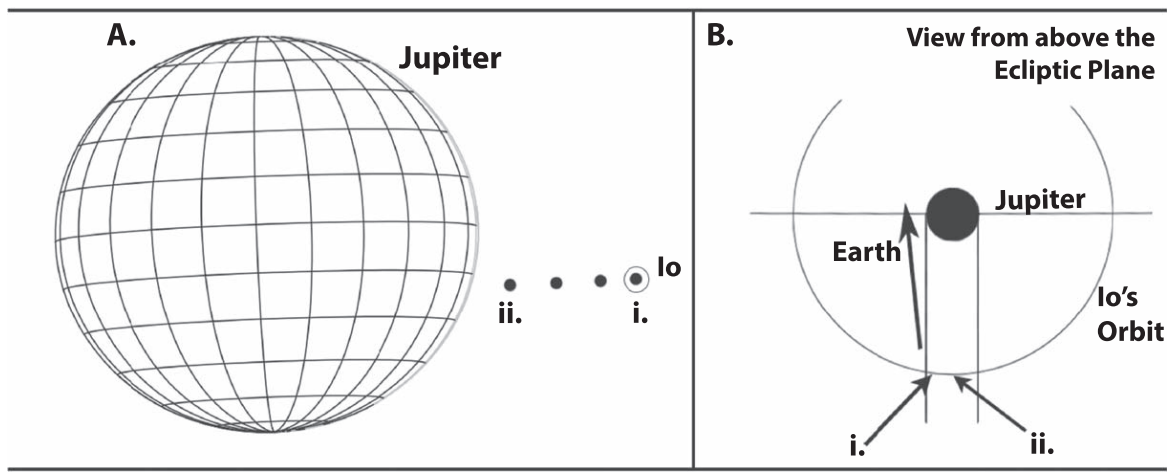

**Figure 2.** Io–Jupiter geometry during the LBT/PEPSI measurement on 2019 April 24. The circle around Io indicates PEPSI's field of view.

While the simultaneous oxygen red line decrease is similar to SO and $SO_2$ millimeter emissions, rising behavior on 2019 April 24 suggests that this trend is coincidental. On both dates, Na D emission sharply decreases immediately after Io enters shadow. This response is rapid, with an e-folding timescale of about 10 minutes after umbral ingress. Like SO and $SO_2$, the Na atmosphere only partially depletes, and emissions appear to reach a new steady state within about 30 minutes. Pre-egress measurements, categorized as "E" in Table 1, are made an hour or more into eclipse. The median Na D brightness in this phase is 1.7 kR, 27% of that measured in the first 10 minutes of shadow. A skeptical reading of the precipitous drop at ingress might be contamination from photon excitation by foreground Na in sunlight. Io's Na cloud structure is very extended (e.g., Grava et al. 2021), and optical sunlight refracting through Jupiter's atmosphere still illuminates Io up to 13 minutes in shadow (Geissler et al. 2004). However, neither changes in the Na Doppler shift nor a pre-egress brightening are observed in our data set, precluding this explanation.

In sunlight, the oxygen red line is known to track the local plasma density of the torus sweeping past Io (Oliversen et al. 2001), and the variability in Figure 5 can be attributed in part to the Io–torus geometry. Figure 6 shows the [O I] red line brightness versus Jovian latitude from the centrifugal equator of the plasma torus. The centrifugal plane is slightly warped because of Jupiter's nondipolar magnetic moments, and this figure uses the surface defined at 5.905 $R_J$ by Phipps & Bagenal (2021). Data points cluster at high latitudes, where Io resides most. Blue curves represent the plasma scale heights in the ribbon (0.71 $R_J$; Phipps et al. 2018) and warm torus (0.90 $R_J$; Moirano et al. 2021), as derived from Juno radio occultations. The Pearson's correlation between red line aurorae and the relative plasma density is 0.53 in the warm torus and 0.55 in the ribbon. Scaling of the blue axis normalizes the warm torus, where Io usually resides, to the 4.8 kR mean emission level, but the denser ribbon region of the torus occasionally extends to Io's radial distance. The torus structure near Io's orbit is complex, as discussed by Thomas et al. (2004), and both the ribbon's radial extent (Schmidt et al. 2018) and the latitudinal plasma scale height (Schneider et al. 1997) depend on the Jovian magnetic longitude. While the actual plasma density sweeping past Io is challenging to predict, Figure 6 confirms that red line emission scales with the upstream plasma environment based on a simplified geometry, where latitude is the primary component. Overall, the scatter and dependence on the torus geometry in Figure 6 both resemble past studies of Io's oxygen aurorae in sunlight (Oliversen et al. 2001; Roth et al. 2014).

Fainter auroral lines can be distinguished only in coadded measurements where temporal information is not fully preserved. Oxygen green line emission is only marginally detected in our data set, and only during times when the red line was relatively bright. Coadding the HIRES measurements, which have the most reliable scattered-light subtraction, shows only a hint of emission near the $\sim$200 R $1\sigma$ level. The night of brightest red line emission, 2020 October 17, is the only $3\sigma$ detection of 5577 Å emission in our data set with $328 \pm 108$ R. This corresponds to a red/green line ratio of $27.6 \pm 10.3$, and the next brightest [O I] data on 2021 October 29 give a red/green of $11.8 \pm 9.1$. These constraints are poor but consistent with the only previously reported red/green [O I] ratio of $18.2 \pm 3.5$ by Bouchez et al. (2000). Their ability to better pin the oxygen line ratio using HIRES stems from the fact that all emissions, including sodium, were several times brighter than we observe here, and discussion of this is deferred to Section 6.

The only new emission we spectroscopically detect at a reasonable significance is potassium 7664.90 Å ($D_2$) and 7698.96 Å ($D_1$), seen in Figure 7. The measured K $D_2 + D_1$ brightness of $507 \pm 93$ R is effectively a lower limit because $D_2$ is blocked by telluric $O_2$ ($b$–$X$)(0–0). Like sodium, the state multiplicity gives an expected $D_2/D_1$ line ratio of 2.0 for electron impact excitations on K atoms. The measured ratio of 1.3 reflects that K $D_2$ is partially absorbed in Earth's atmosphere, while K $D_1$ is transmitted.

### 4. Excitation Rates in Eclipse

Interpretation of these eclipsed emissions is complicated by three factors. First, emissions in shadow could, in principle, originate from multiple mechanisms, including direct electron excitation of an atomic gas, dissociative excitation of molecules, or radiative recombination of molecular ions. Second, inelastic collisions with Io's $SO_2$ efficiently cool the electron population, so electrons that precipitate deep into Io's atmosphere may have insufficient energy to excite the transitions. Third, the oxygen upper states have very long lifetimes that radiate through forbidden transitions, so excited O atoms could be collisionally quenched before emitting light.

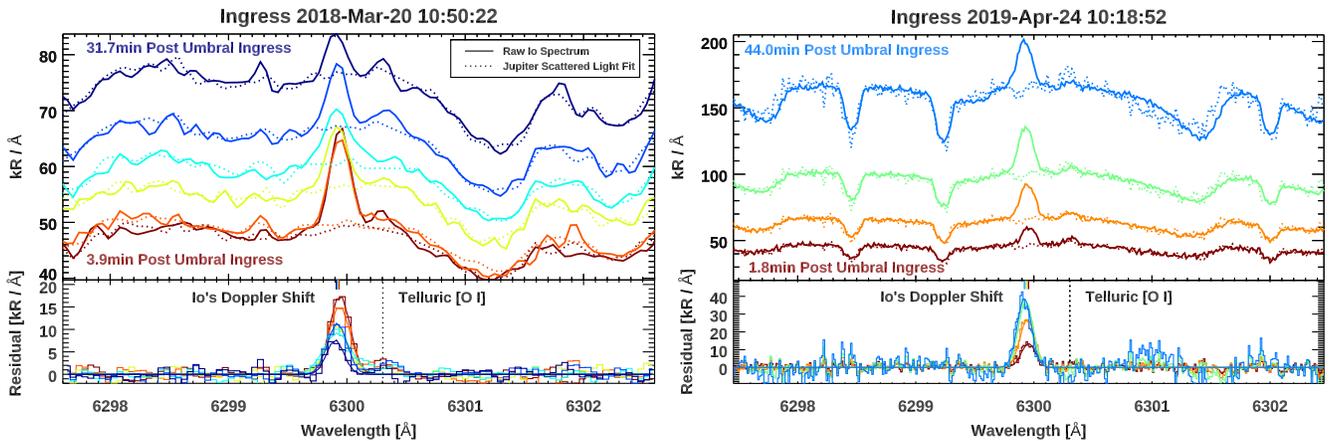

**Figure 3.** Io's [O I] 6300 Å line emission amidst Jupiter scattered light on 2018 March 20 with APO/ARCES (left) and 2019 April 24 with LBT/PEPSI (right). The red-to-blue color scheme shows the elapsed time in shadow. Scattered-light levels increase as Io approaches the Jovian limb, and dotted lines show the fitting of a background spectrum to this component. The lower panels show the residual after scattered-light subtraction. Gaussians are fit to Io's emission line, which Doppler shifts slightly in time. The observing geometry for each date appears in Figures 1 and 2.

Emission produced by electron impact excitations of an atomic gas can be expressed as

$$4\pi I = 10^{-6} N \int_{E_0}^{\infty} \sigma(E) J(E) dE, \quad (1)$$

where $N$ is the atomic column density, $\sigma(E)$ is the excitation cross section by an electron with energy $E$ (in cm$^2$), and $J(E)$ is the differential electron flux (in electrons cm$^{-2}$ s$^{-1}$ eV$^{-1}$). This expression has Rayleigh units, and $E_0$ is the threshold energy corresponding to the upper state of the observed wavelength. Note that the integral in Equation (1), if divided by the electron density, is also conventionally described as a rate coefficient in units of cm$^3$ s$^{-1}$. Figure 8 shows the product of $\sigma(E)J(E)$ for an assumed 3000 electrons cm$^{-3}$ with a 5 eV Maxwell–Boltzmann distribution. This represents the canonical electron density and temperature of the plasma torus at Io's orbit (e.g., Bagenal 1994). Contributions from the trace ∼150 eV hot electron component are negligible, since the $\sigma(E)$ cross section falls off steeply with electron temperature in all cases. Working under the assumption that electron impacts on an atomic gas are the dominant excitation mechanism, integrating these curves from the $E_0$ threshold gives the excitation rate needed to retrieve column density from brightness.

### 4.1. Alkali Excitation

Burger et al. (2001) estimated the sodium column density over Io's sub-Jovian hemisphere at $3 \times 10^{12}$ cm$^{-2}$. For the excitation frequency at the canonical electron flux and temperature of the torus in Figure 8, this Na column would emit 3.5 kR. Other processes can populate Na into the 3p state, augmenting electron impact excitation. The jet feature of energetic sodium escaping Io is produced by NaCl$^+$ recombination above the exobase. Its brightness implies that NaCl$^+$ is produced at rates of $(1-8) \times 10^{26}$ s$^{-1}$ (Wilson & Schneider 1994). It is unknown if recombination or dissociation of NaCl$^+$ populates Na fragments with a 3p or higher state, but if so, the D line contribution would be 400–3000 R integrated over one hemisphere. Of course, Na fragments may also be produced in the ground state or far from Io and outside our aperture, favoring the low brightness range. Electron impact cross sections to excite Na from 3s to 3d states are roughly 5% of those in the 3s to 3p D line transition (Msezane 1988). This radiates the doublet at 8183 and 8194 Å, which cascades ∼175 R (disk average) through the D lines and is below the detection limits. Cross sections for radiative dissociation of NaCl to produce D line emission are unknown, but by neglecting this contribution, 4.1 kR of Na D emission is produced by the aforementioned sources. This theoretical emission is 35% below our measurements immediately following ingress—a fairly good agreement for such a rough calculation.

A Na/K ratio of $10 \pm 3$ (Brown 2001) would produce 610 R of potassium D emission by direct electron impact excitation. Adjusting the measured K D$_2$ brightness to the statistical weight of 2.0 to account for telluric absorption yields a similar value of 663 R. The Geissler et al. (2004) emission model found that potassium was largely responsible for eclipsed flux in the IR1 and IR2 filter bands of the Narrow Angle Camera of Cassini's Imaging Science Subsystem. The relatively bright emission here confirms their finding. Other plausible near-IR sources are O and S. Laboratory studies show that SO$_2$ radiative dissociation yields O I triplets at 7774 and 8446 Å and S I at 9225 Å, where the threshold energy to produce each is ∼25 eV (Ajello et al. 2008). None are evident in the present spectra upon inspection, with upper limits near 250 R confirming that they comprise only a small fraction of the flux measured via spacecraft imaging at Jupiter. The coadded residual spectrum from Keck/HIRES does show two faint peaks of radiative decay from sodium's 3d state, however, wherein the 8195 Å line is stronger, as expected from statistical weights. Since the noise levels are comparable to the expected flux, the features should not be interpreted as a definitive detection, but it is plausible that sodium contributes measurably to the near-IR flux that Cassini observed, and follow-up measurements could confirm this hypothesis.

### 4.2. Oxygen Excitation

The column density of atomic oxygen is less certain than that of sodium. Wolven et al. (2001) estimated $1-2 \times 10^{14}$ cm$^{-2}$ at 2 $R_{Io}$ based on the optically thin 1356 Å brightness there.

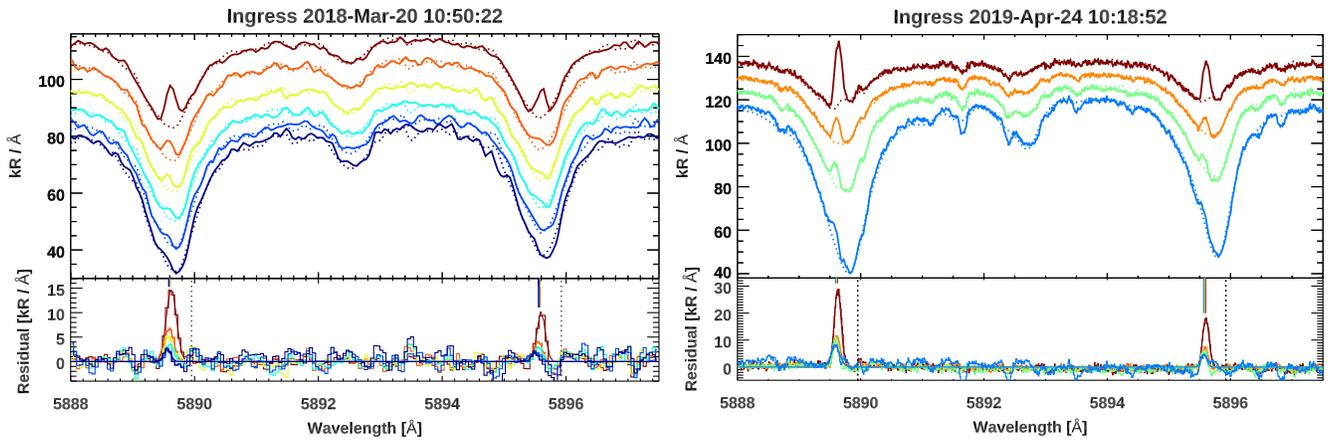

**Figure 4.** Same as Figure 3 but showing the Na D line region with spectra in the top panels consecutively offset by 12 kR Å$^{-1}$ for clarity. No offset is applied to the blue spectrum, which is the last observation taken shortly before Io was occulted by Jupiter's limb.

Smyth & Wong's (2004) models place the O column between $4 \times 10^{14}$ and $1 \times 10^{16}$ cm$^{-2}$. The mean red line emission of 4.8 kR that we measure in eclipse is lower than the 5–10 kR disk-integrated brightness in eclipse measured by HST (Moore et al. 2010) and ground-based measurements in sunlight (Oliversen et al. 2001) and well below the 23.7 kR reported in eclipse from Keck/HIRES (Bouchez et al. 2000). An O column of $6.2 \times 10^{14}$ cm$^{-2}$ is required to emit 4.8 kR via electron impact excitation of atomic oxygen (again assuming 5 eV electron temperature and 3000 cm$^{-3}$ density). However, the long radiative lifetime raises the possibility that the forbidden red line is collisionally deexcited before radiating, in which case the true oxygen column would be higher.

The relative intensity of the red and green oxygen lines provides a constraint, albeit limited, on both the excitation mechanism and the importance of quenching. About 6% of the O($^1S$) radiative decay proceeds though the 2972 Å transition, and decay through 5577 Å produces equivalent red and green emission because of cascade. Accounting for these factors, the excitation frequencies in Figure 8 predict a red/green line ratio of 13.6. Collisional quenching should lower this ratio because, with a lifetime of 134 s, O($^1D$) is collisionally depopulated more rapidly than O($^1S$), which radiatively decays in only 0.8 s. The SO$_2$ collisions quench O($^1D$) at a rate of $2.2 \times 10^{-10}$ cm$^3$ s$^{-1}$ molecule$^{-1}$ (Zhao et al. 2010), and the critical density where SO$_2$ deexcites the atoms before red line decay is $3.4 \times 10^7$ cm$^3$. At the equatorial dayside, models place the corresponding altitude of red line quenching in the 220–400 km range (Moses et al. 2002a; Smyth & Wong 2004). However, O is more globally distributed than SO$_2$. Red line emission appears as a bright limb glow at Io's poles (Geissler et al. 1999), where the SO$_2$ abundance drops by 2 orders of magnitude compared to the equator (e.g., Feaga et al. 2009). Moreover, the fractional coverage of SO$_2$ over Io's disk drops from 30%–35% in sunlight to only 10%–17% in eclipse (de Pater et al. 2020), implying that quenching may be confined to small portions of Io's disk.

On the other hand, inelastic collisions cool torus electrons as they precipitate through Io's atmosphere, which raises the red/green [O I] ratio. Simulations of the Io–torus interaction indicate that electron temperatures exceed the $E_0$ thresholds only under nearly collisionless conditions (Saur et al. 2002; Dols et al. 2012). Io's cold ionosphere is in thermal equilibrium with the atmosphere (∼0.02 eV; Hinson et al. 1998), and electrons precipitating from the torus may have insufficient energy to excite the 4.2 eV green line transition by the time they reach Io's lower atmosphere. The O($^1S$) energy is nearly twice that of the other atomic upper states measured here, so electron cooling could serve to increase the red/green line ratio. Neglecting quenching, column-averaged electron temperatures in the range of 2.1–4.0 eV would correspond to the best-constrained red/green ratio of $18.2 \pm 3.5$ from Bouchez et al. (2000), which is consistent with our results. Electron-SO$_2$ scattering is important in this range, since vibrational modes are resonantly excited at 3.4 eV (Andric et al. 1983), and the cooling of torus electrons may be significant despite the reduced SO$_2$ column in shadow.

Dissociative excitation and radiative recombination could also contribute to oxygen emission. Broadened line widths could be a signature of these reactions, since the molecular bond-breaking can impart excess kinetic energy to the atomic fragments. A survey of sunlit [O I] 6300 Å by Oliversen et al. (2001) reported a positive correlation between brightness and line width, which they attributed to the excess kinetic energy produced by SO$_2$ dissociation. The LBT/PEPSI measurements, which have the highest spectral resolution, could not confirm their finding. The [O I] 6300 Å line widths in eclipse are nearly constant while emission changes, spanning 2.89–3.06 km s$^{-1}$ where the instrumental width has not been deconvolved. This measurably exceeds Na D line widths, which decrease from 2.71 to 2.18 km s$^{-1}$ as the Na emission weakens following ingress.

At present, there are few laboratory data to inform the dissociative excitation and radiative recombination contributions to red line O emission. The cross section for O($^1D$) fragments from dissociative excitation of SO$_2$ by electrons is unknown because experiments require ∼10 nbar pressures to suppress collisional quenching, and achieving this in most laboratory environments is nontrivial. Kedzierski et al. (2000) measured $5 \times 10^{-19}$ cm$^{-2}$ into O($^1S$) with a threshold of 15.9 eV. They also determined a 16.5 eV threshold for the allowed O I 7774 and 8446 Å. At 25 eV, cross sections for three oxygen features are measured, with 7774 Å being 39% and 8446 Å being 22% of the 5577 Å area of $1 \times 10^{-18}$ cm$^{-2}$ (Kedzierski et al. 2000; Ajello et al. 2008). As these fall below the detection limits, our measurements do not preclude the possibility that SO$_2$ dissociative excitation contributes to Io's oxygen aurora, but it is likely a minor contribution, since electron impacts on O alone are adequate to explain line ratios and intensities. Due to inelastic cooling and attenuation in Io's atmosphere, the population of superthermal electrons available to dissociate low-altitude $SO_2$ into $O(^1D)$ fragments is substantially smaller than the $\sim 2\,\mathrm{eV}$ population available to directly excite high-altitude O. Geissler et al. (1999, 2004) also reached this conclusion, consistent with imaging at Jupiter showing red line emission that is bright above Io's poles, more extended, and spatially separated from the equatorial $SO_2$ aurora.

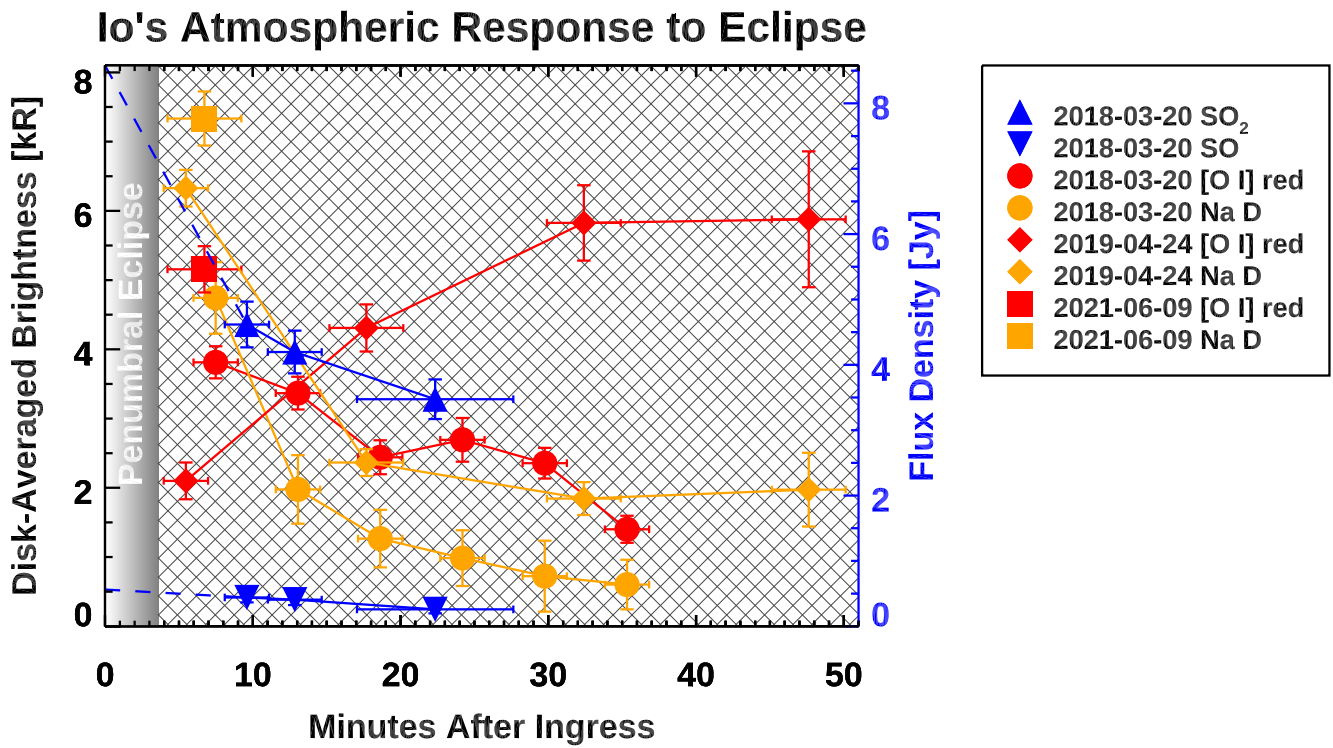

**Figure 5.** Temporal response of Io's atmospheric oxygen (red) and sodium (orange) following ingress. Diamonds were measured using LBT/PEPSI following the 2019 April 24 ingress. Circles show APO/ARCES measurements on 2018 March 20. The concurrent SO and $SO_2$ molecular response reported in millimeter-wavelength ALMA data by de Pater et al. (2020) is shown in blue associated flux density in janskys on the right axis (see their Figure 5). Squares show the only spectrum obtained from APO on 2021 June 9. Data points show summed Na doublet emission, while the [O I] 6364 Å counterpart to 6300 Å is not included, since it falls out of the PEPSI bandpass in one LBT eye. Horizontal error bars show the integration time, and the data points are located mid-exposure.

### 5. Sodium Opacity, Collapse, and Recovery

Sodium D emissions are optically thick above Io's disk. The brightness ratio between the doublet is an opacity diagnostic because the stronger $D_2$ line saturates before $D_1$ (McElroy & Yung 1975). In shadow, the optically thin $D_2/D_1$ ratio is equal to the 2.0 ratio of state multiplicity. Figure 9 shows the measured Na $D_2/D_1$ ratio versus time elapsed in shadow (the K $D_2/D_1$ ratio is unclear, since telluric $O_2$ partially absorbs $D_2$). The uncertainty margins become large as the lines dim with time. The observed ratios are optically thin within their uncertainty margin, with the exception of two data points. An outlying $D_2/D_1$ value below unity does not deviate significantly from 2.0, but the $1.72 \pm 0.11$ ratio immediately following umbral ingress is optically thick at a 99% confidence interval. This suggests that Na D emission evolves from optically thick to thin conditions as the column depletes in eclipse. Consequently, Io's true sodium abundance could change by even more than the brightness in Figure 5 would imply.

Solar irradiance differs at each of the D line wavelengths, so sunlight, unlike plasma, pumps each D line with different rates. The sunlit D line ratio depends slightly on Io's heliocentric Doppler shift, and the photon scattering phase function is isotropic for $D_1$ but not $D_2$. For these reasons, the optically thin Na $D_2/D_1$ ratio is $\sim 1.75$ in sunlight, different than the 2.0 ratio in shadow. Figure 10 shows the sunlit Na D line emission and line ratios following egress. In this case, the uncertainty margins are high near egress, and as Io's heliocentric Doppler shift slowly accelerates sunward, the solar flux for scattering increases, causing brighter D line emission and better opacity constraints. This brightening due to Io's changing Doppler shift is shown in blue (calculated using 2021 October 29 egress geometry). Scattering rates on a fixed sodium column of $5 \times 10^{11}\,\mathrm{cm^{-2}}$ reproduce the flux immediately after egress, but the observed emission increases by more than is predicted from geometric considerations. Concurrently, line ratios also return to optically thick conditions. The bottom panel of Figure 10 shows $D_2/D_1$ near the optically thin ratio following egress (binned every 15 minutes in black) but returning to optically thick conditions after some 45 minutes in sunlight. As with ingress measurements, this again suggests that the actual changes in Na column abundance are more extreme than the change in brightness alone would suggest.

Figure 11 shows a curve-of-growth model with estimates of the changing sodium column density on 2021 October 29, the most comprehensive observing date following an egress. Colored curves in both panels are interpolated from Brown & Yung (1976) to a gas temperature of 1600 K (Burger et al. 2001), and their assumed solar irradiance is scaled to Io's instantaneous heliocentric distance and Doppler shift. The colored curves are not overly sensitive to this temperature choice, since opacity scales as the inverse square root of temperature. At 1600 K, the Na $D_2$ line center reaches unity optical depth at $2.3 \times 10^{11}$ atoms cm$^{-2}$, above which the column density versus brightness relationship is no longer linear. This represents a global treatment of highly localized atmospheric structure, and the two metrics for the column density—brightness and line ratio—do not agree precisely. Be that as it may, both indicate a slow sodium rise with time. The apparent order-of-magnitude growth should be interpreted with caution, since it is difficult to accurately constrain the column immediately following egress. Both metrics suggest an increase from very roughly $3 \times 10^{11}$ to $10^{12}$ atoms cm$^{-2}$ over the course of 2.5 hr. This range of Na growth is more moderate than the factor of >3.7 precipitous drop at ingress, despite the much longer timescale over which the egress response is observed.

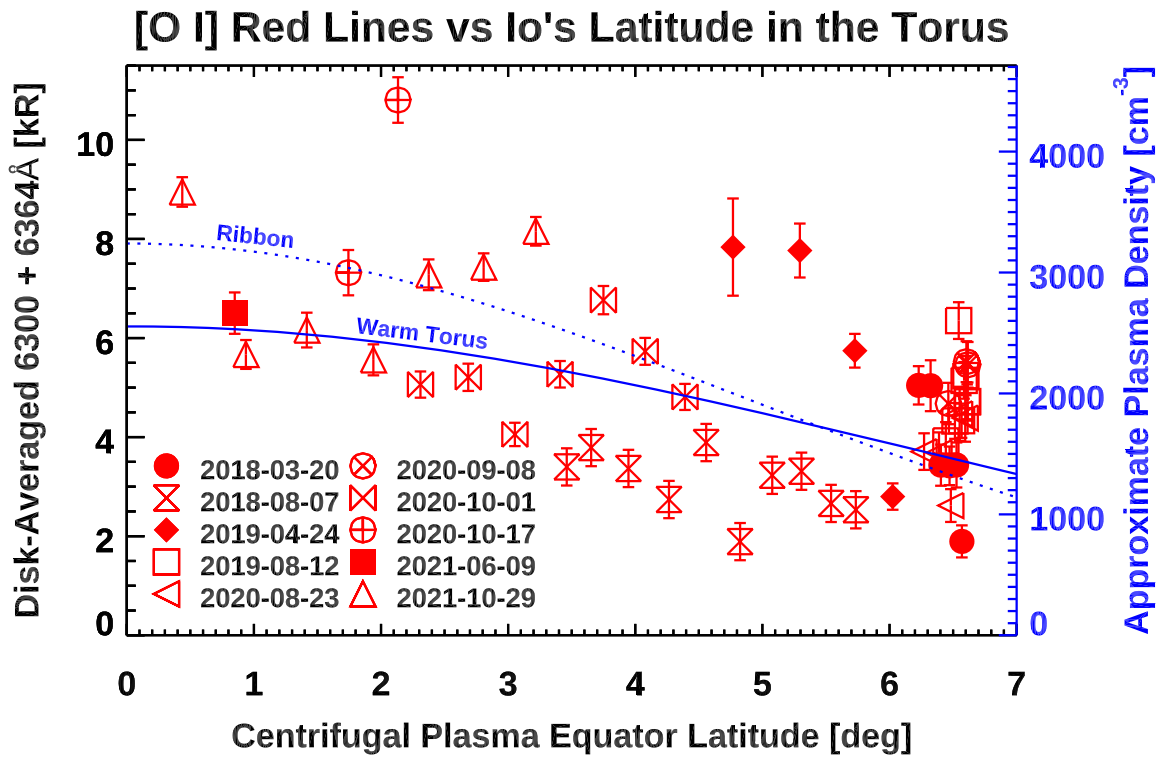

**Figure 6.** Effects of the changing torus latitude on Io's disk-averaged emission from the [O I] red line doublet. Filled symbols represent observations following ingress, and open symbols represent observations preceding egress. The blue curve shows the approximate plasma density of the ribbon and warm torus regions based on Juno radio occultations.

### 5.1. Interpretation of the Sodium Eclipse Phase Response

#### 5.1.1. Potential Coupling to the $SO_2$ Response

Behavioral similarity between $SO_2$ and Na in Figure 5 is suggestive of an interrelated process in their eclipse response. Both collapse with comparable timescales of approximately 10 minutes. Both only partially collapse: Na toward <27% and $SO_2$ toward 30%–50% of the preeclipse value. There are also plausible mechanisms that could couple Na and $SO_2$. Atmospheric sputtering of Na could be suppressed if Na-bearing molecules are buried during $SO_2$ condensation. The $SO_2$ winds, which may circulate Na from volcanic sources to broader regions of Io's atmosphere, also stagnate in eclipse (Walker et al. 2012). This explanation raises the question: why would sodium exhibit a strong response, but not oxygen? Unlike Na, quenching may confine forbidden O emission to a narrow range of atmospheric density. Lowering both the O and $SO_2$ abundance only pushes the [O I] emitting region closer to Io's surface—a small change in the overall cross section that Io's atmosphere presents to the torus. Thus, if quenching is ubiquitous, the nonresponse of red and green O emission is not necessarily in conflict with a strong Na eclipse response that stems from some form of coupling with $SO_2$.

#### 5.1.2. Potential Interruption of NaCl Photodissociation

Photochemical models suggest that roughly 99.6% of Na is produced by photodissociation: NaCl + $h\nu \rightarrow$ Na + Cl (Moses et al. 2002b). Grava et al. (2014) posed the interruption of this Na production channel as an explanation for the long atmospheric Na recovery time at egress. The most rapid photochemical loss for Na is thought to be a three-body reaction, $SO_2$+Na+M $\rightarrow$ $NaSO_2$+M, at a rate of 18 days (Moses et al. 2002b), and the partial collapse of $SO_2$ would slow this even further. However, Na also efficiently escapes from NaCl that dissociates near the exobase. Near the poles and far from plumes, the local atmosphere over broad regions of Io's surface may constitute a surface-bound exosphere, particularly in eclipse and presumably on the nightside. The excess energy of the ionic bond and atmospheric sputtering imparts a velocity for Na to escape Io into the "banana cloud" at rates of $(1-9) \times 10^{26}$ atoms s$^{-1}$, and the total Na escape rate including ion loss is $(3-25) \times 10^{26}$ atoms s$^{-1}$ (Wilson et al. 2002). Consequently, if photolytic pathways supplying Na were turned off in shadow, then this total escape rate would exhaust the entire $3 \times 10^{12}$ cm$^{-2}$ sub-Jovian Na column in only 4–35 minutes, which is broadly consistent with the response timescale in Figure 5 at ingress.

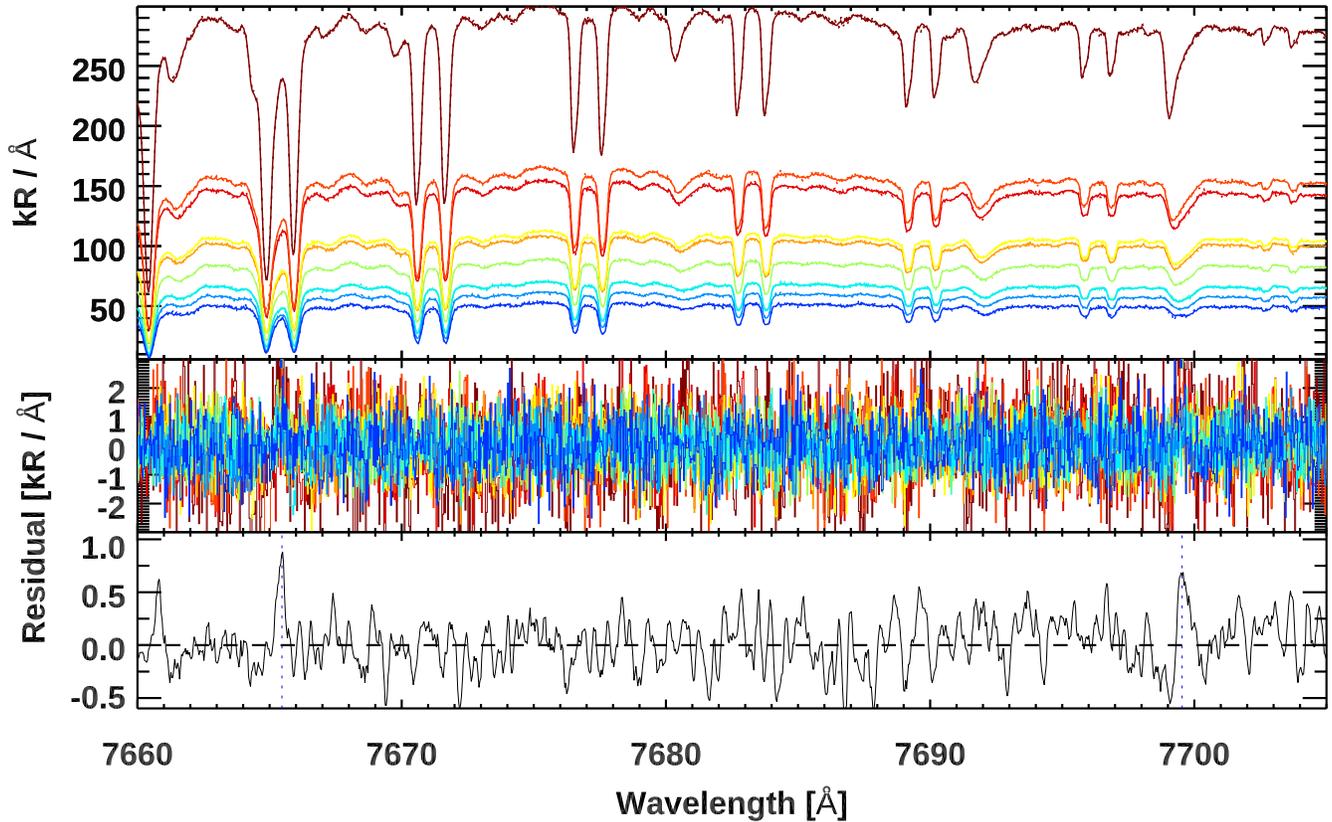

**Figure 7.** Potassium D line emissions seen just preceding egress with Keck/HIRES on 2018 August 7. The top panel shows nine raw spectra that have the Jovian scattered light subtracted off in the middle panel. The bottom panel shows the median of these frames after coaligning for Io's changing Doppler shift. The stronger $D_2$ line at 7665 Å is partially absorbed by telluric $O_2$, making the $D_2/D_1$ line ratio lower than the statistical value of 2.0.

The photodissociation lifetime of NaCl in sunlight is not firmly constrained. Near-UV dissociation cross sections reported by Davidovits & Brodhead (1967) at 1123–1223 K are several times higher than those of Silver et al. (1986) measured at 300 K, and the NaCl lifetime at Io corresponds to 26 minutes (Schaefer & Fegley 2005) or 3.2 hr (Moses et al. 2002b), depending on which cross section is used. Quantum chemistry calculations are consistent with the shorter lifetime (Valiev et al. 2020) and show that the temperature dependence of the cross section alone cannot account for the differences reported in past literature (Pezzella et al. 2021).

The changing sodium emissions at ingress and egress offer an important clue to the underlying photochemistry: the time constant for Na loss at ingress (∼10 minutes) is much more rapid than the recovery time constant at egress (∼2 hr). Io's NaCl atmosphere is quite stable (Roth et al. 2020), and in steady state, the production of Na must balance its high loss rates. If NaCl is sourced from direct volcanic output (Lellouch et al. 2003; Redwing et al. 2022), and NaCl photodissociation produces nearly all of Io's Na (Moses et al. 2002b), then the loss and recovery timescales should equate. Such a disparate timescale at ingress and egress suggests another process. Simulations by Walker et al. (2012) showed that the $SO_2$ collapse at ingress should occur more quickly than the recovery at egress because the high thermal inertia of the $SO_2$ frost itself causes it to warm and sublimate more slowly. However, recent ALMA-based measurements show rapid and similar timescales for atmospheric $SO_2$ collapse and recovery (de Pater et al. 2020). Thus, the different Na loss and recovery timescales that we observe disfavor the two above hypotheses of coupling to $SO_2$ or photodissociation of NaCl.

### 5.1.3. Potential Interruption of Photochemical Pathways for $NaCl^+$

Ion recombination, which is understood from observations of the "jet" feature to produce Na in high abundance (Schneider et al. 1991), could result in different Na fading and recovery rates. For $Na^+$ or $NaCl^+$ produced at rate $P$, recombination changes the Na volume density as

$$\frac{d[n_{Na}]}{dt} \approx -\frac{d[n_{ion}]}{dt} = \alpha[n_e][n_{ion}] - P_{ion}. \qquad (2)$$

Photochemical models suggest that $Na^+$ and $K^+$ are predominant in Io's ionosphere as the terminal ions (Moses et al. 2002b; Summers & Strobel 1996). Still, the recombination rate coefficient for sodium cations, $\alpha = 2.76\,(T/300)^{-0.68} \times 10^{-12}$ cm$^3$ s$^{-1}$ (McElroy et al. 2013), is so inefficient as to be negligible. The $NaCl^+$ recombination rate coefficient is unknown, but a reasonable estimate is $\alpha \sim 2 \times 10^{-7}$ cm$^3$ s$^{-1}$ (Larsson & Orel 2008). Thus, if photochemistry causes the $P_{NaCl+}$, $[n_e]$ or $[n_{NaCl+}]$ terms in Equation (2) to be high at ingress and relatively low at egress, then ion recombination could plausibly explain the observed Na temporal response. Escape can again explain the overall magnitude of the response; ionic pathways for Na escape are comparable to or faster than the neutral pathway (Wilson et al. 2002), and if $P \sim 0$ in shadow, the sub-Jovian Na atmosphere could escape with roughly the requisite 10 minute e-folding timescale.

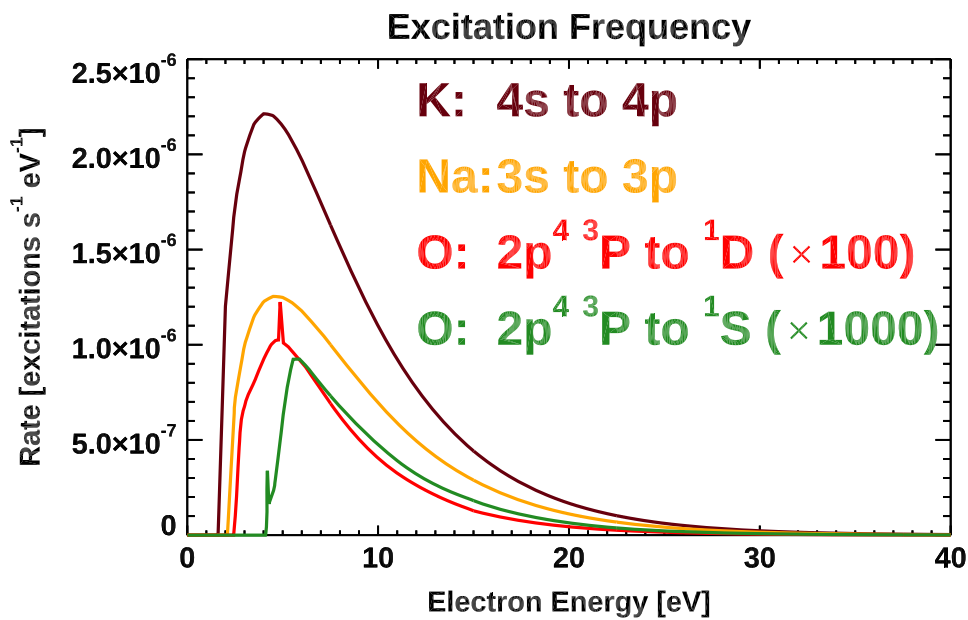

**Figure 8.** Electron impact excitation frequencies on an atomic gas for all observed electronic transitions. Alkali impact cross sections are taken from Stone & Kim (2004). Oxygen impact cross sections are taken from Barklem (2007).

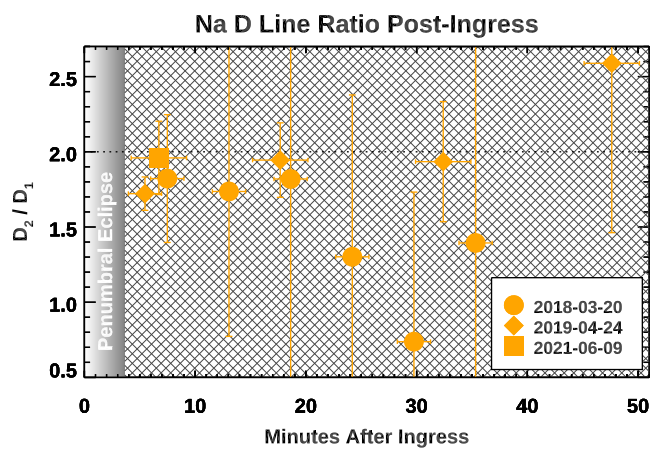

**Figure 9.** Sodium D line ratios following ingress. A ratio of 2.0 is expected for optically thin emission, and only the ratio nearest ingress on 2019 April 24 deviates from this value with appreciable significance ($\sigma = 2.5$; cf. Figure 4).

It is uncertain how the ionosphere might evolve during the ~2 hr long eclipse phase. Radio occultations offer the best constraints on the ionospheric density, $[n_e]$, but their geometry is limited to paths within 10° of the terminator. Density weakly depends on solar zenith angle in this range (Hinson et al. 1998; see their Figure 13), suggesting that photons compete with the torus in ion production, and precipitating electrons may become too cold to ionize at low altitudes where photons could be more important. Photochemical models alone can sustain a peak ion density of $\sim 10^5$ cm$^{-3}$ (Moses et al. 2002b; Summers & Strobel 1996), but 3D simulations suggest that the torus ionization better explains Galileo flyby measurements (Saur et al. 2002, 1999). The Saur & Strobel (2004) model demonstrates nonlinear relations between $[n_e]$ and column density during the eclipse phase, affecting auroral brightness. This coupling may also result in a nonlinear production rate of the ionosphere with time in eclipse.

At ingress, the sub-Jovian hemisphere should resemble the dayside ionosphere, which is very extended, with a 200–400 km scale height and peak $[n_e] \sim 6 \times 10^4$ cm$^{-3}$ (Kliore et al. 1975; Hinson et al. 1998). The NaCl$^+$ lifetime against recombination is inversely proportional to the plasma density, and immediately following ingress, $1/\alpha[n_e] = 1.4$ minutes at the peak of the ionosphere, sufficiently fast to explain the ingress response if there is minimal NaCl$^+$ production in shadow. Even if ~2 hr in shadow dramatically reduced Io's ionosphere into equilibrium with the background torus density, the lifetime against recombination grows to only ~30 minutes, so it is unlikely that $[n_e]$ changes alone can explain the Na response timescales at both ingress and egress.

Insufficient ionospheric change during eclipse leads us to conclude that NaCl$^+$ must be sensitive to sunlight through the $P$ and/or $[n_{NaCl+}]$ terms. Solar production of NaCl$^+$ is likely indirect, since photoionization of NaCl is thought to be negligible compared to photodissociation (Heays et al. 2017), so a different reaction probably produces NaCl$^+$ from NaCl. The photochemistry we propose requires only that photo-ionization of some species dominates the production pathway to NaCl$^+$. This could be a multistep reaction where NaCl$^+$ could be efficiently supplied by charge transfer with molecular ions like SO$^+$, SO$_2^+$, and S$_2^+$ (Johnson 1994; Moses et al. 2002b). Ion cyclotron waves near Io indicate that a substantial portion of SO$_2$ is ionized before dissociating, and the inferred SO$_2^+$ production rate of $8 \times 10^{26}$ s$^{-1}$ is more than sufficient to supply NaCl$^+$ at rates observed in the Na jet (Huddleston et al. 1998). This also implies that the source of SO$_2^+$ is at nearly collisionless altitudes, since the fresh ring-shaped velocity distribution perpendicular to the magnetic field is needed to initialize wave growth, and collisions disallow such anisotropy.

In summary, two criteria are needed for our hypothesis to be viable: (1) that NaCl$^+$ is predominant within the chemical pathways producing Io's atomic Na and (2) that the predominant pathway for NaCl$^+$ production involves solar photochemistry. If both of these hold, then solar-produced NaCl$^+$ recombination satisfies all aspects of the observed Na response to eclipse. Io's sodium column of $\sim 3 \times 10^{12}$ cm$^{-2}$ can be lost within 10 minutes of ingress, assuming a recombination timescale of <2 minutes in the peak ionosphere and transport losses above Io's exobase in the range of $5 \times 10^{25}$ (Burger et al. 1999) to $3 \times 10^{26}$ s$^{-1}$ (Wilson et al. 2002). Recovery timescales of a few hours measured in this work and Grava et al. (2014) are then attributed to a confluence of the bulk atmospheric recovery time, the time for multistep photochemistry to create NaCl$^+$, and its slower recombination within a weakened ionosphere at egress.

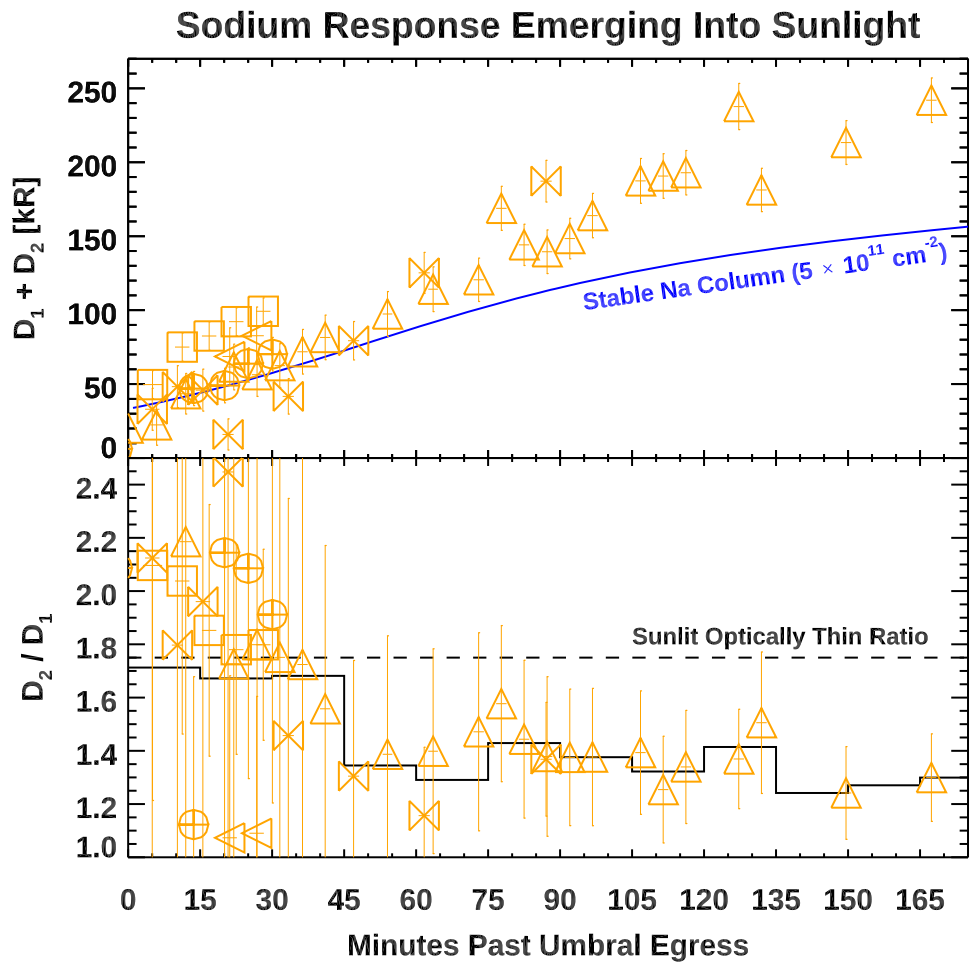

**Figure 10.** Sodium D emission (top) and line ratios (bottom) as a function of duration in sunlight postegress. Symbols indicate different observing dates following the Figure 6 legend. The top panel shows the summed $D_1 + D_2$ emission. Io's heliocentric Doppler shift causes increased sunlight to resonantly scatter in the atoms, so a static column, shown in blue, will brighten in time. A constant column density cannot reproduce the data points, however, suggesting that Io's Na density is rebuilding after the eclipse phase. The optically thin line ratio of 1.75 is shown in the bottom panel, where 15 minute binned data (black) become optically thick after 45 minutes, confirming growth in the Na abundance.

## 6. Discussion

Emissions from multiple lines of O, Na, and K are observed from Io in eclipse with ground-based telescopes. All emissions are broadly consistent with the brightness levels expected from electron impact on atomic gases. Other excitation mechanisms, namely, radiative dissociation and recombination, are poorly constrained empirically, but both produce one photon per reaction, which is inefficient considering that one electron can yield many photons. Line ratios suggest that inelastic cooling during electron precipitation through Io's atmosphere may influence its oxygen aurorae, and that collisions are overall ineffective in quenching [O I] emissions, at least in shadow.

All five emission lines of Na and O in the present data set are a factor of 4–5 lower than Bouchez et al. (2000) reported, including measurements with the same HIRES instrument. The disk-averaged brightness of these lines is also less than previous eclipse measurements from Galileo (Geissler et al. 1999) and HST (Moore et al. 2010). The line ratios remain consistent with previous findings, however, suggesting that long-term changes in the torus's plasma density may be responsible. Optical surveys of torus S$^+$ emissions, which scale as the plasma density squared, encompass this dynamic range (Schmidt et al. 2018). However, the brightness we report could also be underestimated due to the small aperture size that is sampled in the observations—only marginally larger than the angular size of Io itself. This limitation is inherent to most high-resolution optical spectrographs, as their design is optimized for point sources. Atmospheric seeing and pointing inaccuracies could displace light outside the apertures used here. Moreover, some studies show [O I] 6300 Å concentrated in Io's wake (Moore et al. 2010), which lies 0.″6 from the aperture center and only 0.″20–0.″55 from the periphery in our instruments.

Io's optical oxygen emissions are nonresponsive to eclipse. The [O I] red line brightness approximately tracks the changing relative plasma density with Io's location in the torus, but Na D emission does not show a similar correlation, even in the subset of measurements taken well after ingress. Sodium appears highly responsive to eclipse, and its flux quickly drops by a factor of 3.7 in shadow, consistent with the sunlit measurements by Grava et al. (2014) postegress. The actual change in Io's sodium column density is greater still, as emissions traverse a boundary between optically thick and thin conditions. It would be a misnomer to construe this sodium response as atmospheric collapse; during eclipse, Io's $SO_2$ atmosphere is removed via condensation to the surface, while Io's Na is removed at the top of the atmosphere via escape.

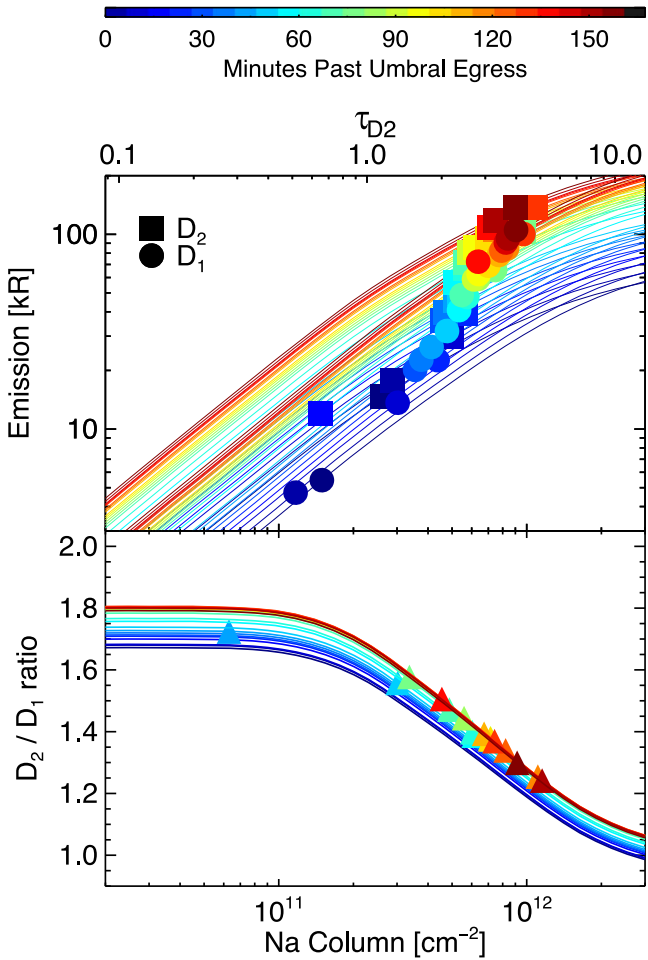

**Figure 11.** Curve-of-growth models for sodium $D_1$ and $D_2$ emission (top) and their line ratio (bottom) as a function of time on 29 October 2021. Color coding indicates the time past egress. Measured emissions and line ratios are plotted vs. their column density inferred from each metric. The opacity treatment here derives from Brown & Yung (1976) and is interpolated to an assumed gas temperature of 1600 K. Incident sunlight at the $D_1$ and $D_2$ wavelengths is changing in time due to Io's heliocentric Doppler shift. Hence, there is a vertical spread in the top panel curve-of-growth lines; i.e., emission changes with time, even if the column density is fixed.

The atmospheric sodium behavior surrounding eclipse is attributed to $NaCl^+$ recombination and removal of this ion-borne pathway for Na production in shadow. Alternate interpretations are possible, but this hypothesis best explains both an ∼10 minute e-folding collapse at ingress and a 2 hr e-folding recovery timescale at egress, where Na production is hampered by the long photolysis time to produce $NaCl^+$ and slower recombination in a weakened ionosphere. The removal of energetic atoms in the absence of $NaCl^+$ recombination is also consistent with observed narrowing of the Na D line widths following ingress, lending confidence to this interpretation. An observational test for our hypothesis is whether Io's sodium jet feature disappears in shadow; its absence near egress would confirm that the Na production pathway via $NaCl^+$ requires sunlight. As in Grava et al. (2014), the full recovery time for Na to return to preingress values exceeds the duration of our measurements. When comparing, note that their study characterized the Na response at altitudes of 3 $R_{Io}$ and higher to avoid the optically thick region over Io's disk. Transport to these altitudes could take hours for $NaCl^+$ recombination below the exobase, and Na produced above the exobase could quickly escape. Together, these studies reveal a scenario in which ion chemistry significantly alters atmospheric composition, a phenomenon unique to Io.

Previous potassium measurements have been limited to the cloud far from Io, and the electron-excited K in Figure 7 is not only the first data to constrain the K abundance above Io's disk but also the first confirmation of such emission in a planetary atmosphere. Based on the excitation rates in Section 4 and assuming $[n_e] \sim 2000$ cm$^{-3}$ per Figure 5, the O:Na:K column abundance ratio is 1200:2.8:1 with a K column of $4.9 \times 10^{11}$ atoms cm$^{-2}$ during the 2018 August 7 measurement. The true O abundance is at least slightly higher, depending on the fraction of [O I] red line emission that is quenched by noncondensed gas. Taking the red line–derived column as a lower limit, O/Na is >425 near egress. At ingress, this ratio is >135 and could be nearly stoichiometric with the disk-integrated $SO_2$/NaCl of 53–91 reported in sunlight (Roth et al. 2020). Presently, there is no means of verifying whether Na and K respond similarly to eclipse, so their ratio may also deviate with Io's phase. This sole measurement near egress yields Na/K of 2.8, in good agreement with the 1–7 NaCl/KCl ratio determined by Redwing et al. (2022). If K were stable throughout eclipse, however, the results would be more consistent with the sunlit Na/K ratio of $10 \pm 3$ that Brown (2001) reported beyond 10 $R_{Io}$. At this distance, differences in Na and K transport and lifetime against ionization become important. Observed line profiles suggest that dissociation of K molecular ions is weaker than Na (Thomas 1996; Schmidt 2022), and electron impacts preferentially remove K (Carlson et al. 1975).

## 7. Summary and Future Work

Io's atomic aurorae are characterized at optical wavelengths using spatially unresolved spectra from ground-based telescopes. The main scientific results can be summarized as follows.

1. The peak efficiency for electron impact excitation of major transitions in O, Na, and K atoms also coincides with the core electron temperature of the Io plasma torus; hence, this is determined to be the primary emission mechanism for Io's aurorae at optical wavelengths.
2. Io's red line oxygen aurora shows a dependence on the satellite's location relative to the plasma torus and appears unaffected by its passage though Jupiter's shadow. Assuming a torus influx of 3000 electrons cm$^{-3}$ at 5 eV, a stable O column of $6.2 \times 10^{14}$ cm$^{-2}$ is inferred over the sub-Jovian hemisphere.
3. Red-to-green oxygen emission ratios are broadly consistent with the sole previous measurement (Bouchez et al. 2000) and higher than expected from O electron impact. Cooling of torus electrons as they precipitate through the atmosphere is a plausible explanation, despite the reduced molecular atmosphere in shadow.

4. The Na exhibits radically different depletion and recovery timescales during the eclipse phase. The e-folding timescale is 10 minutes and 2 hr at ingress and egress, respectively.
5. The Na $D_2/D_1$ line ratio also changes surrounding eclipse, corresponding to a transition from optically thick to thin emissions in shadow.
6. The temporal behavior of Na emissions is interpreted as a budget imbalance in its production and loss channels surrounding eclipse. Interruption of the $NaCl^+$ photochemical production pathway in shadow and rapid loss via escape offer the most viable explanation of the observed characteristics.
7. The above result implies that sunlight is required to sustain $NaCl^+$ at Io; however, $NaCl^+$ cannot be produced by solar photoionization directly, implying that efficient charge transfer must occur between another species, perhaps $SO_2^+$ or $SO^+$.
8. A positive correlation between Na brightness and line width is observed in shadow, qualitatively consistent with Na fragments being imparted excess kinetic energy by molecular dissociation. No such relationship was present in O.
9. Auroral emissions from the D line doublet allow a first estimate of the potassium column above Io's disk: $4.9 \times 10^{11}$ cm$^{-2}$ averaged over the sub-Jovian hemisphere. This measurement was made just prior to egress, and the temporal response of auroral K during eclipse remains unconstrained.

In future work, comparing long-slit high-resolution spectroscopy during the pre- and postegress phases would inform whether Io's sodium jet feature is sensitive to eclipse, as we predict. While we argue that $NaCl^+$ is a predominant source of Na, this pathway only presents a loss of order 1% to the NaCl volcanic outgassing, as Lellouch et al. (2003) pointed out. New ALMA data analysis appears consistent with the expectation that NaCl abundance is unchanged by the absence of photochemistry in shadow (Redwing et al. 2022). We also propose that Io's sodium emits in the 8183 and 8195 Å doublet at brightness levels of a few hundred Rayleighs. If follow-up observations confirm this assertion, then the near-IR to D line ratio would offer a potential constraint on the local electron temperature without the complications of collisional quenching that are inherent to forbidden O aurorae. Additional HIRES observations surrounding Io's ingress are currently scheduled and may resolve ambiguity on the potassium response to eclipse. Observational constraints on the response of chlorine are more challenging because, while HST/COS has the sensitivity to measure temporal changes, the FUV Cl lines are probably optically thick.

Support for the observations and analysis in this study was provided by the National Aeronautics and Space Administration (NASA) Solar System Observations program, 80NSSC22K0954 and 80NSSC21K1138. M.S. acknowledges support from the Massachusetts Space Grant Consortium. K.d. K. acknowledges support from NASA through grant No. HST-GO-15425.002-A from the Space Telescope Science Institute, which is operated by AURA, Inc., under NASA contract NAS 5-26555. J.M. gratefully acknowledges support from the National Science Foundation program AST-2108416. Constraints on the plasma conditions at Io were posed via Juno data, and support for that analysis derives from NASA program 80NSSC19K0818. L.M. was supported by grant 80NSSC20K1045 issued through the NASA Solar System Workings program. We thank Jean McKeever of Yale University for her help with reduction of the Apache Point echelle data and Candace Gray and Russet McMillan for their assistance with those observations. The LBT is an international collaboration among institutions in the United States, Italy, and Germany. The LBT Corporation partners are the University of Arizona on behalf of the Arizona Board of Regents; Istituto Nazionale di Astrofisica, Italy; LBT Beteiligungsgesellschaft, Germany, representing the Max Planck Society, the Leibniz Institute for Astrophysics Potsdam, and Heidelberg University; and the Ohio State University, representing OSU, the University of Notre Dame, the University of Minnesota, and the University of Virginia. Some of the data presented herein were obtained at the W. M. Keck Observatory, which is operated as a scientific partnership among the California Institute of Technology, the University of California, and the National Aeronautics and Space Administration. The Observatory was made possible by the generous financial support of the W. M. Keck Foundation. The authors wish to recognize and acknowledge the very significant cultural role and reverence that the summit of Maunakea has always had within the indigenous Hawaiian community. We are most fortunate to have the opportunity to conduct observations from this mountain.

### ORCID iDs

Carl Schmidt ● https://orcid.org/0000-0002-6917-3458
Katherine de Kleer ● https://orcid.org/0000-0002-9068-3428
Nick Schneider ● https://orcid.org/0000-0001-6720-5519
Imke de Pater ● https://orcid.org/0000-0002-4278-3168
Phillip H. Phipps ● https://orcid.org/0000-0002-4323-4400
Albert Conrad ● https://orcid.org/0000-0003-2872-0061
Paul Withers ● https://orcid.org/0000-0003-3084-4581
John Spencer ● https://orcid.org/0000-0003-4452-8109
Jeff Morgenthaler ● https://orcid.org/0000-0003-3716-3455
Ilya Ilyin ● https://orcid.org/0000-0002-0551-046X
Klaus Strassmeier ● https://orcid.org/0000-0002-6192-6494
Christian Veillet ● https://orcid.org/0000-0003-0272-0418
John Hill ● https://orcid.org/0000-0003-2484-3670
Mike Brown ● https://orcid.org/0000-0002-8255-0545

### References

Ajello, J. M., Aguilar, A., Mangina, R. S., et al. 2008, JGRE, 113, E03002
Andric, L., Cadez, I., Hall, R. I., et al. 1983, JPhB, 16, 1837
Bagenal, F. 1994, JGR, 99, 11043
Barklem, P. S. 2007, A&A, 462, 781
Bouchez, A. H., Brown, M. E., & Schneider, N. M. 2000, Icar, 148, 316
Brown, M. E. 2001, Icar, 151, 190
Brown, R. A., & Yung, Y. L. 1976, Jupiter: Studies of the Interior, Atmosphere, Magnetosphere and Satellites (Tucson, AZ: Univ. Arizona Press), 1102
Burger, M. H., Schneider, N. M., de Pater, I., et al. 2001, ApJ, 563, 1063
Burger, M. H., Schneider, N. M., & Wilson, J. K. 1999, GeoRL, 26, 3333
Carlson, R. W., Matson, D. L., & Johnson, T. V. 1975, GeoRL, 2, 469
Clarke, J. T., Ajello, J., Luhmann, J., et al. 1994, JGR, 99, 8387
Davidovits, P., & Brodhead, D. C. 1967, JChPh, 46, 2968
de Kleer, K., & Brown, M. E. 2018, AJ, 156, 167
de Pater, I., Luszcz-Cook, S., Rojo, P., et al. 2020, PSJ, 1, 60

Dols, V., Delamere, P. A., Bagenal, F., et al. 2012, JGRE, 117, E10010
Feaga, L. M., McGrath, M., & Feldman, P. D. 2009, Icar, 201, 570
Geissler, P., McEwen, A., Porco, C., et al. 2004, Icar, 172, 127
Geissler, P. E., McEwen, A. S., Ip, W., et al. 1999, Sci, 285, 870
Geissler, P. E., Smyth, W. H., McEwen, A. S., et al. 2001, JGR, 106, 26137
Grava, C., Cassidy, T. A., Schneider, N. M., et al. 2021, AJ, 162, 190
Grava, C., Schneider, N. M., Leblanc, F., et al. 2014, JGRE, 119, 404
Heays, A. N., Bosman, A. D., & Van Dishoeck, E. F. 2017, A&A, 602, 105
Hinson, D. P., Kliore, A. J., Flasar, F. M., et al. 1998, JGR, 103, 29343
Huddleston, D. E., Strangeway, R. J., Warnecke, J., et al. 1998, JGR, 103, 19887
Johnson, R. E. 1994, Icar, 111, 65
Kedzierski, W., Malone, C., & McConkey, J. W. 2000, CaJPh, 78, 617
Kliore, A. J., Fjeldbo, G., Seidel, B. L., et al. 1975, Icar, 24, 407
Larsson, M., & Orel, A. E. 2008, Dissociative Recombination of Molecular Ions (Cambridge: Cambridge Univ. Press)
Lellouch, E., Paubert, G., Moses, J. I., et al. 2003, Natur, 421, 45
McElroy, D., Walsh, C., Markwick, A. J., et al. 2013, A&A, 550, A36
McElroy, M. B., & Yung, Y. L. 1975, ApJ, 196, 227
Moirano, A., Gomez Casajus, L., Zannoni, M., et al. 2021, JGRA, 126, e29190
Moore, C., Miki, K., Goldstein, D. B., et al. 2010, Icar, 207, 810
Moses, J. I., Zolotov, M. Y., & Fegley, B. 2002a, Icar, 156, 76
Moses, J. I., Zolotov, M. Y., & Fegley, B. 2002b, Icar, 156, 107
Msezane, A. Z. 1988, Phys Rev A, 37, 1787
Oliversen, R. J., Scherb, F., Smyth, W. H., et al. 2001, JGR, 106, 26183
Pezzella, M., Yurchenko, S. N., & Tennyson, J. 2021, PCCP, 23, 16390
Phipps, P., & Bagenal, F. 2021, JGRA, 126, e2020JA028713
Phipps, P. H., Withers, P., Buccino, D. R., et al. 2018, JGRA, 123, 6207
Redwing, E., de Pater, I., Luszcz-Cook, et al. 2022, PSJ, 3, 238
Retherford, K. D., Moos, H. W., & Strobel, D. F. 2003, JGRA, 108, 1333
Retherford, K. D., Spencer, J. R., Stern, S. A., et al. 2007, Sci, 318, 237
Roth, L., Boissier, J., Moullet, A., et al. 2020, Icar, 350, 113925
Roth, L., Saur, J., Retherford, K. D., et al. 2011, Icar, 214, 495
Roth, L., Saur, J., Retherford, K. D., et al. 2014, Icar, 228, 386
Saur, J., Neubauer, F. M., Strobel, D. F., et al. 1999, JGR, 104, 25105
Saur, J., Neubauer, F. M., Strobel, D. F., et al. 2002, JGRA, 107, 1422
Saur, J., & Strobel, D. F. 2004, Icar, 171, 411
Schaefer, L., & Fegley, B. 2005, Icar, 173, 454
Schmidt, C. 2022, FrASS, 9, 801873
Schmidt, C., Schneider, N., Leblanc, F., et al. 2018, JGRA, 123, 5610
Schneider, N. M., Taylor, M. H., Crary, F. J., et al. 1997, JGR, 102, 19823
Schneider, N. M., Trauger, J. T., Wilson, J. K., et al. 1991, Sci, 253, 1394
Silver, J. A., Worsnop, D. R., Freedman, A., et al. 1986, JChPh, 84, 4378
Smyth, W. H., & Wong, M. C. 2004, Icar, 171, 171
Spencer, J. R., Stern, S. A., Cheng, A. F., et al. 2007, Sci, 318, 240
Stone, P. M., & Kim, Y. K. 2004, J Res Natl Inst Stand Technol, 109, 505
Strassmeier, K. G., Ilyin, I., Järvinen, A., et al. 2015, AN, 336, 324
Strassmeier, K. G., Ilyin, I., Weber, M., et al. 2018, Proc. SPIE, 10702, 1070212
Summers, M. E., & Strobel, D. F. 1996, Icar, 120, 290
Thomas, N. 1996, A&A, 313, 306
Thomas, N., Bagenal, F., Hill, T., et al. 2004, in Jupiter: The Planet, Satellites and Magnetosphere, ed. F. Bagenal, T. Dowling, & W. McKinnon (Cambridge: Cambridge Univ. Press), 561
Tsang, C. C. C., Spencer, J. R., & Jessup, K. L. 2015, Icar, 248, 243
Tsang, C. C. C., Spencer, J. R., Lellouch, E., et al. 2016, JGRE, 121, 1400
Valiev, R. R., Berezhnoy, A. A., Gritsenko, I. S., et al. 2020, A&A, 633, 39
Vogt, S. S., Allen, S. L., Bigelow, B. C., et al. 1994, Proc. SPIE, 2198, 362
Walker, A. C., Moore, C. H., Goldstein, D. B., et al. 2012, Icar, 220, 225
Wang, S., Hildebrand, R. H., Hobbs, L. M., et al. 2003, Proc. SPIE, 4841, 1145
Wilson, J. K., Mendillo, M., Baumgardner, J., et al. 2002, Icar, 157, 476
Wilson, J. K., & Schneider, N. M. 1994, Icar, 111, 31
Wolven, B. C., Moos, H. W., Retherford, K. D., et al. 2001, JGR, 106, 26155
Zhao, Z., Laine, P. L., Nicovich, J. M., et al. 2010, PNAS, 107, 6610